\documentclass{article} 
\usepackage{graphicx}

\newcommand{\bspec}{\mbox{$\beta$-spectrum}}
\newcommand{\belec}{\mbox{$\beta$-electron}}
\newcommand{\bdec}{\mbox{$\beta$-decay}}
\newcommand{\nuone}{\mbox{$\nu_{\mathrm 1}$}}
\newcommand{\nutwo}{\mbox{$\nu_{\mathrm 2}$}}
\newcommand{\nuthree}{\mbox{$\nu_{\mathrm 3}$}}
\newcommand{\nufour}{\mbox{$\nu_\mathrm{4}$}}
\newcommand{\mnufour}{\mbox{$m(\nu_\mathrm{4})$}}
\newcommand{\mtwonufour}{\mbox{$m^2(\nu_\mathrm{4})$}}

\newcommand{\nue}{\mbox{$\nu_{\mathrm e}$}}
\newcommand{\numu}{\mbox{$\nu_{\mu}$}}
\newcommand{\nutau}{\mbox{$\nu_{\tau}$}}
\newcommand{\nus}{\mbox{$\nu_\mathrm{s}$}}

\newcommand{\nui}{\mbox{$\nu_{\mathrm i}$}}

\newcommand{\mtwonui}{\mbox{$m^2(\nu_{\mathrm i})$}}
\newcommand{\mtwonulight}{\mbox{$m^2(\nu_\mathrm{light})$}}

\newcommand{\mnui}{\mbox{$m(\nu_{\mathrm i})$}}

\newcommand{\mnulight}{\mbox{$m(\nu_\mathrm{light})$}}

\newcommand{\mnue}{\mbox{$m(\nu_{\mathrm e} )$}}

\newcommand{\mtwonue}{\mbox{$m^2(\nu_{\mathrm e} )$}}

\newcommand{\mbb}{\mbox{$m_{\beta\beta}$}}

\newcommand{\mhad}{\mbox{$M_{\rm nucl}$}}

\newcommand{\mtwohad}{\mbox{$|M^2_{\rm nucl}|$}}
\newcommand{\ezero}{\mbox{$E_0$}}

\newcommand{\vj}{\mbox{$V_{\rm j}$}}
\newcommand{\pj}{\mbox{$P_{\rm j}$}}
\newcommand{\evtwo}{\mbox{eV$^2$}}

\newcommand{\be}{\begin{equation}}
\newcommand{\ee}{\end{equation}}
\newcommand{\bea}{\begin{eqnarray}}
\newcommand{\eea}{\end{eqnarray}}

\newcommand{\etal}{\mbox{\it et al.}}
\begin{document}

\title{Limit on sterile neutrino contribution from the Mainz Neutrino Mass Experiment}

\author{Christine Kraus$^1$,  
Andrej Singer$^2$\footnote{present address: Deutsches Elektronen-Synchrotron DESY, Notkestr. 85, 22607 Hamburg, Germany}, 
Kathrin Valerius$^2$\footnote{present address:  Erlangen Centre for Astroparticle Physics, Erwin-Rommel-Str. 1, 91058 Erlangen, Germany}
and Christian Weinheimer$^2$\footnote{corresponding author, email: weinheimer@uni-muenster.de}\\
\it \footnotesize $^1$ Laurentian University - 935 Ramsay Lake Road, Sudbury P3E 2C6, Canada\\
\it \footnotesize $^2$ Institute of Nuclear Physics, Westf\"alische Wilhelms-Universit\"at M\"unster,\\
\it \footnotesize Wilhelm-Klemm-Str. 9, 48149 M\"unster, Germany
}

\maketitle

\begin{abstract}
The recent analysis of the normalization of reactor antineutrino data, the calibration data of solar neutrino experiments using gallium targets,
and the results from the neutrino oscillation experiment MiniBooNE suggest the existence of a fourth light neutrino mass state with a mass
of {$\cal O$}(eV), which contributes to the electron neutrino with a sizable mixing angle. 
Since we know from measurements of the width of the Z$^0$ resonance that there are only three active neutrinos, a fourth neutrino should be sterile (i.e., interact only via gravity). 
The corresponding fourth neutrino mass state should be visible as an additional kink in \bdec\ spectra.
In this work the phase II data of the Mainz Neutrino Mass Experiment have been analyzed searching for a possible contribution
of a fourth light neutrino mass state. No signature of such a fourth mass state has been found and limits on the mass and the mixing of this
fourth mass state are derived.
\end{abstract}

\section{Introduction}
\label{intro}
Experiments with atmospheric, solar, reactor and accelerator neutrinos gave compelling evidence that neutrinos from one flavor state can be detected in another flavor state after some flight distance. This well-established phenomenon {\it neutrino oscillation} 
is usually explained by neutrino mixing: {  Firstly,} the three flavor neutrino states \nue , \numu\ and \nutau\ are superpositions of three
neutrino mass eigenstates \nuone , \nutwo\ and \nuthree\ connected by a unitary  $3 \times 3$ mixing matrix $U$. Secondly, neutrino oscillations require
that the three neutrino mass states differ in masses, i.e. at least two neutrino mass states \nui\ possess non-zero masses.
Neutrino oscillation experiments yielded the three mixing angles $\theta_\mathrm{23}$ ($\sin^2\theta_\mathrm{23} = (3.86^{+0.24}_{-0.21})\cdot 10^{-1}$), $\theta_\mathrm{12}$ ($\sin^2\theta_\mathrm{12} = (3.07^{+0.18}_{-0.16})\cdot 10^{-1}$) and recently $\theta_\mathrm{13}$ ($\sin^2\theta_\mathrm{13} = (2.41^{+0.25}_{-0.25})\cdot 10^{-2}$), as well as the two splittings between squared neutrino masses 
$\Delta m^2_\mathrm{12} = m^2(\nutwo ) - m^2(\nuone ) = (7.54 ^{+0.26}_{-0.22}) \cdot 10^{-5}$~\evtwo\ and
$ | \Delta m^2_\mathrm{23} | = | m^2(\nuthree ) - m^2(\nutwo )|  = (2.43 ^{+0.16}_{-0.10}) \cdot 10^{-3}$~\evtwo\ (all values quoted after \cite{fogli12}, using conventional units with $c = 1$ and $\hbar = 1$)\footnote{The values here are given under the assumption of the normal neutrino mass hierarchy, i.e. $m(\nu_3)>m(\nu_2) > m(\nu_1)$, which slightly differ from the results for the inverted hierarchy $m(\nu_2)>m(\nu_1) > m(\nu_3)$.}.

Nearly all these oscillation experiments -- performed with neutrinos with very different energies, different flavors, different flight distances and with/with\-out
matter effects -- can be described by three neutrino mass and three neutrino flavor states connected by a unitary $3 \times 3$ matrix. 
Yet, there is an increasing number of hints that this picture is not complete: There is the request for at least one additional scale of neutrino squared mass splittings
$\Delta m^2_\mathrm{ij} =\cal O$(eV) from the so called {\it reactor neutrino anomaly} \cite{Men11}, 
the normalization of gallium solar neutrino experiments \cite{gallex95,gallex10,sage06,sage09}, and from
the accelerator neutrino experiments LSND \cite{lsnd98} and MiniBooNE \cite{miniboone07}. Although cosmology 
gave hints that the number of neutrino degrees of freedom is rather four than three, introducing an eV mass scale is not trivial \cite{hamann11}.
Since we know from the LEP studies of the Z$^0$ pole that the number of active neutrinos coupling to the {W$^\pm$ and the Z$^0$ bosons
is $2.9840 \pm 0.0082$ \cite{lep06} the fourth neutrino has to be sterile. By neutrino mixing the fourth neutrino state will become visible in neutrino oscillation and direct neutrino mass experiments \cite{giunti11}. The existence of sterile neutrinos is quite natural, because most theories describing non-zero neutrino masses exhibit
right-handed and therefore sterile neutrinos. What is less natural is the eV-scale discussed here.
A summary of the physics and searches for sterile neutrinos can be found in a recent white paper \cite{sterilenu_whitepaper12}.
 
Our paper is organized as follows: In section 2 we discuss the influence of a fourth sterile neutrino on the spectrum of an allowed \bdec . 
The Mainz Neutrino Mass Experiment is described briefly in section 3. In section 4 we present the result of a sterile neutrino search in the 
phase II data of the Mainz Neutrino Mass Experiment before we give a conclusion and an outlook in section 5.

\section{Neutrino mass signature in $\beta$-decay}
The energy spectrum $\dot N(E)$ of the \belec s of an allowed \bdec\
is given by \cite{weinheimer03,otten08}:
\bea  
  \dot N(E)  =  & & \frac{G_\mathrm{F}^2 \cdot \cos^2\Theta_{\rm C}}{2 \pi^3}
              \cdot \mtwohad \cdot F(E ,Z') \cdot  (E + m) \cdot \sqrt{(E + m)^2 - m^2} \nonumber \\
             & \cdot & \sum_{\rm i,j} |U^2_{\rm ei}| \cdot P_{\rm j} \cdot (\ezero - \vj - E) 
                                                       \cdot \sqrt{(\ezero - \vj - E)^2 - \mtwonui}  \nonumber \\
                      & \cdot & \Theta(\ezero - \vj - E - \mnui). \label{eq:betaspec} 
\eea
Here $E$ and $m$ denote the kinetic energy and mass of the electron, $G_\mathrm{F}$ and $\Theta_\mathrm{C}$ Fermi's constant and the Cabibbo angle, 
\mhad\ the nuclear matrix element of the \bdec , $F(E,Z')$ the Fermi function describing the Coulomb interaction of the outgoing electron 
with the remaining daughter nucleus of charge $Z'$, \pj\ the probabilities to find the daughter ion after the \bdec\ in an electronic or rotational-vibrational excitation with excitation
energy \vj , and \ezero\ the maximum possible kinetic energy of the \belec\ in case of $m(\nu_\mathrm{i}) = 0$, which is the $Q$-value of the decay minus the
recoil energy of the daughter \cite{otten08}. $\Theta$ is the Heaviside function.
 
Assuming a fourth sterile neutrino \nus\ (or even a larger number of sterile neutrinos) requires to increase the number of neutrino flavor and mass states as well as the dimensions of the mixing matrix $U$:
\be
  \left( \begin{array}{c}
    \nue \\
    \numu \\
    \nutau \\
    \nus \\
\end{array}  \right)
 =   \left( \begin{array}{llll}
   U_\mathrm{e1}       &U_\mathrm{e2}       & U_\mathrm{e3}      & U_\mathrm{e4}\\ 
   U_\mathrm{\mu 1}  & U_\mathrm{\mu 2} & U_\mathrm{\mu 3} &  U_\mathrm{\mu 4}\\ 
   U_\mathrm{\tau 1}  &U_\mathrm{\tau 2} & U_\mathrm{\tau 3}  & U_\mathrm{\tau 4}\\ 
  U_\mathrm{s1}        & U_\mathrm{s2}      & U_\mathrm{s3}       & U_\mathrm{s4}\\ 
\end{array}   \right)  \cdot
  \left( \begin{array}{c}
    \nuone \\
    \nutwo\\
    \nuthree \\
    \nufour\\
\end{array}   \right). \label{eq:fourbyfour_mixing}
\ee 
The sum over the neutrino mass states in equation (\ref{eq:betaspec}) will then run from $i=1$ to $i=4$. Now we introduce the following simplification:
The three neutrino states \nuone , \nutwo\ and \nuthree\ are assumed to have about the same mass $\mnulight \approx m(\nuone ) \approx m(\nutwo ) \approx m(\nuthree )$. 
This assumption\footnote{We could even allow small differences between the three light neutrino masses $m(\nuone )$, $m(\nutwo )$,
$m(\nuthree )$ and expand the $j$-th component of the 
\bspec\ to first order in $\mtwonui/(\ezero - V_\mathrm{j} - E)^2$ \cite{otten08}.
Then we can define the {\it electron neutrino mass squared} as average over all light mass eigenstates contributing to the electron neutrino: 
$\mtwonue := \sum_\mathrm{i=1}^3 |U_\mathrm{ei}|^2 \, \mtwonui$,
which would correspond to the mass of the light
neutrino mass state  $\mtwonue \approx \mtwonulight$.} is supported by the small differences
between the squared neutrino masses found in neutrino oscillations (see section 1). 
We can now sum up the first three terms $|U_\mathrm{ei}^2|$ in equation (\ref{eq:betaspec}) and describe it by a single mixing angle $\vartheta$:
\be
  \sum_\mathrm{i=1}^3 |U_\mathrm{ei}^2| =: \cos^2(\vartheta), \qquad  |U^2_\mathrm{e4}| =:  \sin^2(\vartheta).
  \label{eq:two_nu_mixing}
\ee 
Equation (\ref{eq:betaspec}) then simplifies to 
{
\bea
  \dot N(E)  =  & & \frac{G_\mathrm{F}^2 \cdot \cos^2\Theta_{\rm C}}{2 \pi^3}
              \cdot \mtwohad \cdot F(E ,Z') \cdot  (E + m) \cdot \sqrt{(E + m)^2 - m^2} \nonumber \\
             & \cdot & \sum_{\rm j} P_{\rm j} \cdot (\ezero - \vj - E) \nonumber \\
              & \cdot & \biggl( \cos^2(\vartheta) \cdot \sqrt{(\ezero - \vj - E)^2 - \mtwonulight} \cdot \Theta(\ezero - \vj - E - \mnulight ) \nonumber \\
              & \quad & +   \sin^2(\vartheta) \cdot \sqrt{(\ezero - \vj - E)^2 - \mtwonufour} \cdot \Theta(\ezero - \vj - E - m(\nufour ))
              \biggr).   \label{eq:betaspec_nusterile}
\eea
}

From equation (\ref{eq:betaspec_nusterile}) it is obvious that the endpoint region of a \bspec\ is the most sensitive region to search for a contribution of a sterile
neutrino with a mass $m(\nufour )= {\cal O}$(1~eV). Therefore, tritium and $^\mathrm{187}$Re are the $\beta$-emitters of choice due to their endpoint energies of $\ezero = 18.57$~keV and $\ezero = 2.47$~keV, respectively, which are the two lowest known $\beta$-endpoint energies.
Figure \ref{fig:betaspec_with_two_nustates} shows a \bspec\ near its endpoint \ezero\ for an arbitrarily chosen contribution of a fourth sterile neutrino mass state.

\begin{figure}[t!]
   \centerline{\includegraphics[angle=0,width=0.6\textwidth]{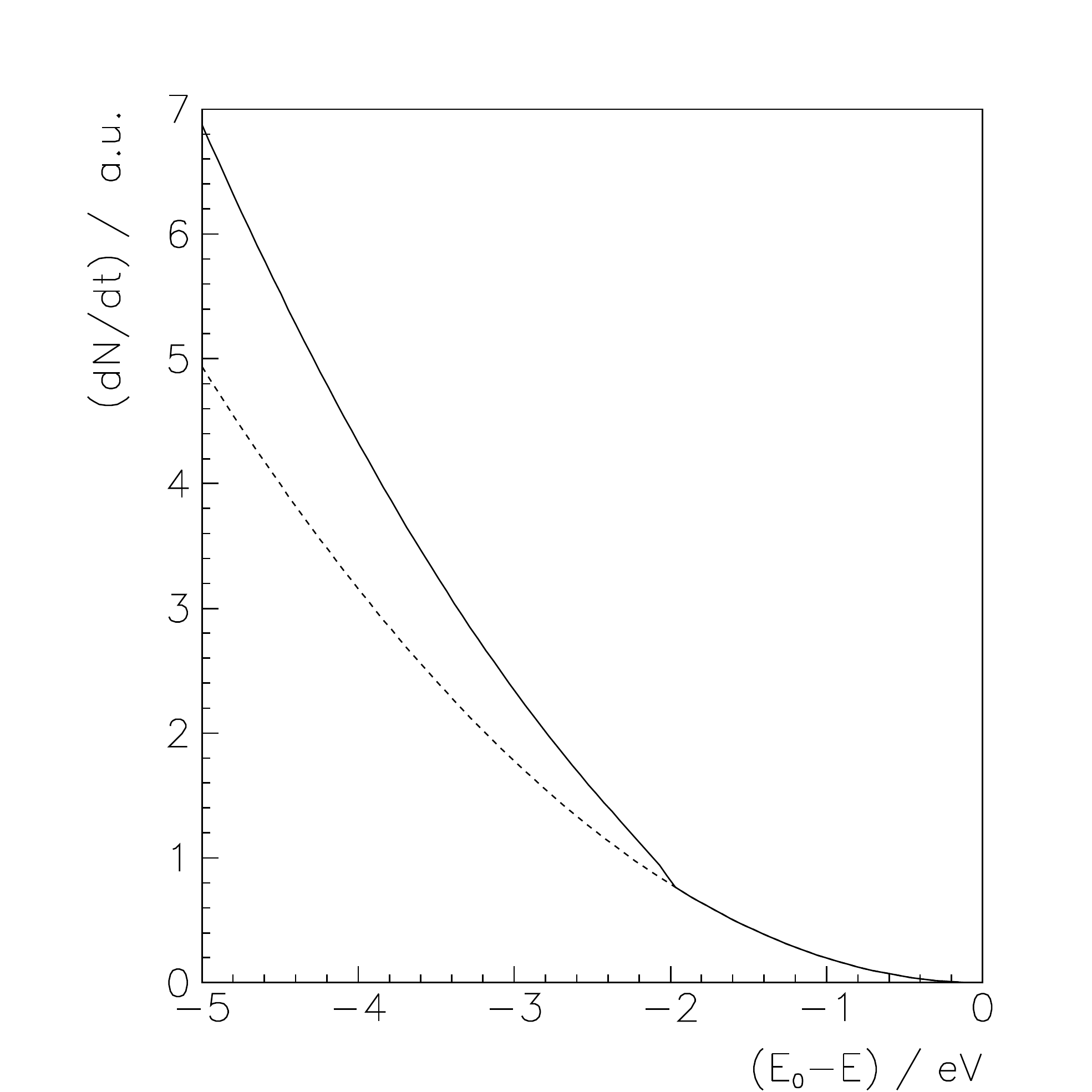}}
  \caption{Allowed \bspec\ near the endpoint \ezero\ with an admixture of a heavy neutrino with $m(\nufour ) = 2$~eV, $\sin^2(\vartheta) = 0.3$. The dashed line shows the \bspec\ with the light neutrino state $\mnulight = 0$~eV only.} 
  \label{fig:betaspec_with_two_nustates}
\end{figure}

\clearpage

\section{Phase II of the Mainz Neutrino Mass Experiment}
The Mainz Neutrino Mass Experiment was investigating the endpoint region of the tritium \bspec\ from 1991 to 2001 to search for a non-zero neutrino mass
\cite{weinheimer99,kraus05}. In the following we will only describe and discuss the data of the Mainz phase II (1998--2001) after the upgrade, which took place from 1995 until 1997.

The Mainz Neutrino Mass Experiment used
an integrating electrostatic retardation spectrometer with magnetic guiding and collimating field of MAC-E-Filter type \cite{pic92a}. 
This spectrometer type combines a large accepted solid angle with a high energy resolution. The retardation potential was created by a system of 27 cylindrical
electrodes, which were installed within an ultrahigh vacuum vessel of 1~m diameter and 3~m length.
The \bspec\ was scanned over the last 200~eV below the endpoint \ezero\ by setting about 40 different retarding voltages and counting the corresponding
number of transmitted \belec s.
To eject stored electrons, which could cause background, from the spectrometer,
HF pulses on one of the electrodes were applied for about 3~s every 20~s between the measurements at a constant retarding voltage.
A system of 5 superconducting solenoids provided
the magnetic guiding field for the \belec s from the tritium source in the first solenoid through a magnetic chicane through the spectrometer to a silicon detector. 
The tritium source consisted of a thin film of molecular tritium, which was quench-condensed 
onto a cold graphite substrate and kept at a temperature below 2~K. By laser ellipsometry the film thickness was determined to be typically 150 monolayers.
The magnetic chicane eliminated source-correlated background. 
The low temperature below 2~K 
avoided the roughening transition of the homogeneously condensed tritium films with time
\cite{fleischmann00}. 
Figure \ref{fig:mainz_newsetup} illustrates the Mainz setup.

\begin{figure}[t!]
    \centerline{\includegraphics[angle=0,width=0.8\textwidth]{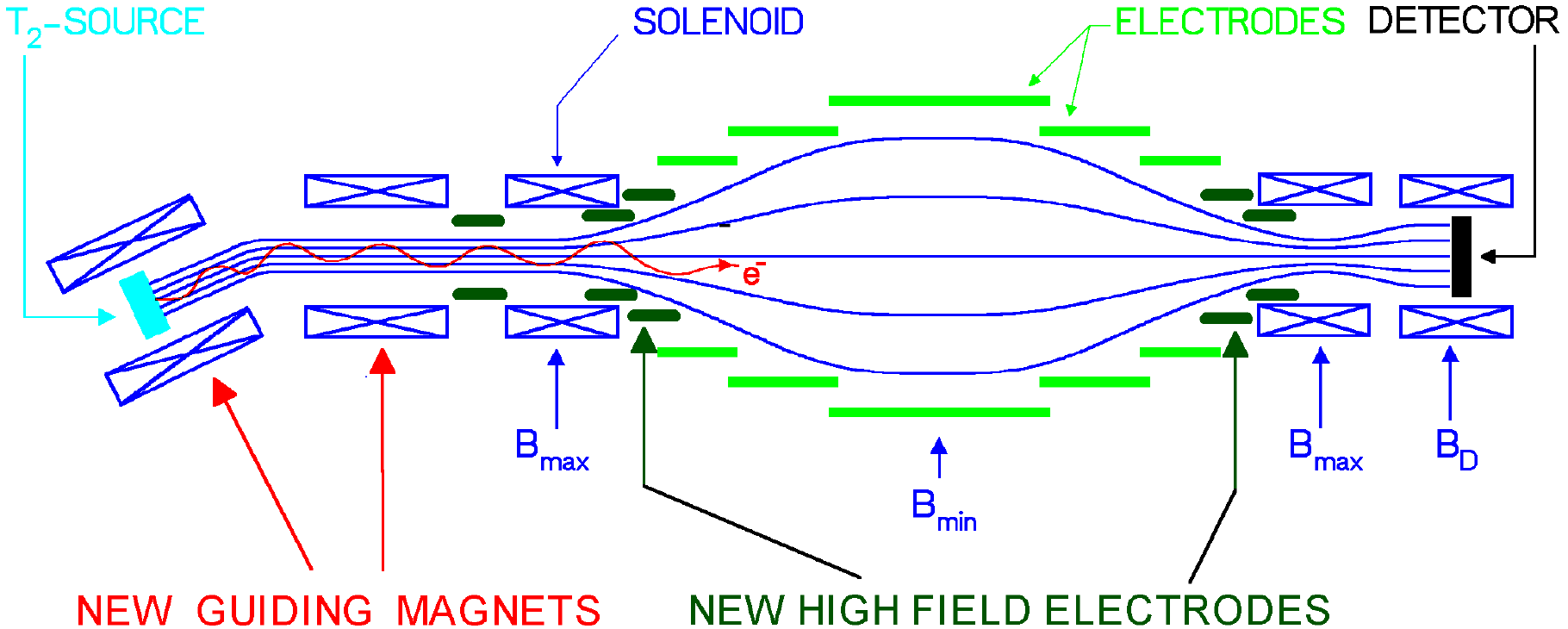}}
  \caption{Illustration of the Mainz Neutrino Mass Experiment after its upgrade for {\it phase II}. 
    The outer diameter of the spectrometer amounted to 1~m, the distance
    from source to detector was 6~m. See text for details.}
  \label{fig:mainz_newsetup}
\end{figure}

\begin{figure}[tb]
\centerline{\includegraphics[width=0.75\textwidth]{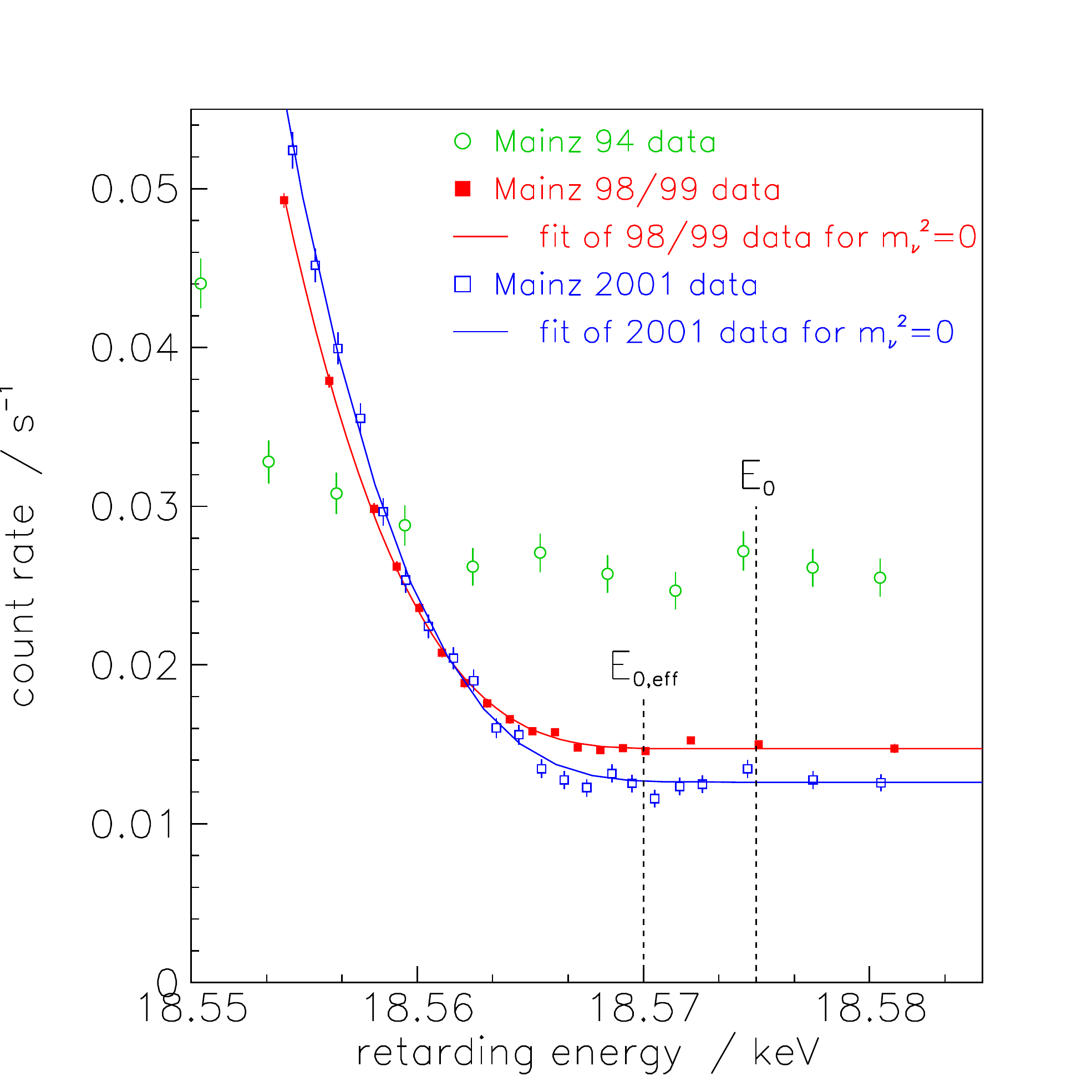}}
\caption{Averaged count rate of the Mainz 1998/1999 data
(filled red squares) with fit for \mnue=0 (red line) 
and of the 2001 data (open blue squares) with fit for \mnue=0 (blue line) 
in comparison with previous Mainz data
from 1994 (open green circles)  
as a function of the retarding energy
 near the endpoint \ezero\  and effective endpoint $E_{0,{\rm eff}}$. The latter takes into account
the width of the response function of the setup and 
the mean rotation-vibration excitation energy of the electronic 
ground state of the $\rm (^3HeT)^+$ daughter molecule.}
\label{fig:mainz_data}
\end{figure}

In 1998, 1999 and 2001 in total 6 runs of about one month duration each were taken under good and well-controlled experimental conditions. These data were analyzed with regard to the neutrino mass \mnue\ \cite{kraus05}.
The main systematic uncertainties of the Mainz experiment
were the inelastic scattering 
of \belec s within the tritium film,
the excitation of neighbor molecules due to sudden change of the nuclear charge
during \bdec , 
and the self-charging of the tritium film as a consequence of its radioactivity. 
As a result of detailed investigations in Mainz  
\cite{aseev00,barth_erice97,bornschein03,kraus05}
-- mostly by dedicated experiments -- 
the systematic corrections became
much better understood and their uncertainties were reduced
significantly.
The high-statistics Mainz phase II data (1998--2001) allowed the first
determination of the probability of the neighbor excitation, which was found 
to occur in $(5.0 \pm 1.6 \pm 2.2)~\%$ of all \bdec s \cite{kraus05}, in good agreement
with the theoretical expectation \cite{kolos88}.

The analysis of the last 70~eV below the endpoint of the phase II data 
gave no indication for a non-zero neutrino mass (see also figure \ref{fig:mainz_data}) \cite{kraus05}. 
The result for the squared neutrino mass,
\begin{equation} \label{eq:mainz_phaseII}
\mtwonue = (-0.6 \pm 2.2 \pm 2.1)~ \mathrm{eV^2}, 
\end{equation}
corresponds  -- using the Feldman-Cousins method  \cite{feldman98} -- to an upper limit  of
\begin{equation} \label{eq:mainz_phaseII_limit}
\mnue < 2.3~ \mathrm{eV} \quad {\rm (95~\%~C.L.).}
\end{equation}

\clearpage 

\section{Analysis of Mainz phase II data with respect to sterile neutrino contribution}
\label{sec:analysis}

We now analyze the six runs of the Mainz phase II data with regard to a possible contribution of a sterile neutrino.
In the former Mainz phase II analysis the following four fit parameters were determined by a fit  to the data: the electron neutrino mass squared \mtwonue , 
the endpoint energy \ezero, the normalizing amplitude $a$ and a constant background rate $b$. 

For our sterile neutrino analysis we assume that the light neutrino mass can be neglected: $\mnulight = \mnue = 0$ compared to the mass of the heavy sterile neutrino \mnufour .
This is not only justified by the neutrino mass limit shown in equation (\ref{eq:mainz_phaseII_limit}) but also from a similar limit of the neutrino mass experiment at Troitsk \cite{aseev11} and from even more stringent neutrino mass limits from cosmology (e.g. \cite{abazajian11}).   
To describe the \bspec\ including a fourth sterile neutrino mass state 
we use equation (\ref{eq:betaspec_nusterile}). 
Hence, we have now five fit parameters instead of the four fit parameters of the previous 
neutrino mass analysis: the squared mass of the heavy sterile neutrino \mtwonufour, its contribution by mixing  $\sin^2 \vartheta$, 
the endpoint energy \ezero, the normalizing amplitude $a$ and a constant background rate $b$.
 
In all other respects we do exactly the same as for the Mainz phase II analysis
\cite{kraus05}. 
This includes using the same data sets of the six runs, which comprise
the measured count rates as a function of the retarding voltage $U$, as well as applying
the same so-called response function $T'(E,U)$ of the Mainz apparatus.
Our fit function $F(U)$ should describe the expected count rate as function of the retarding voltage $U$ of the spectrometer. To obtain this fit function the \bspec\ 
$\dot N(E)$ including a fourth, sterile neutrino (\ref{eq:betaspec_nusterile}) is convolved 
with the response function $T'(E,U)$ of the apparatus
and a constant background rate $b$ is added:
\be
F(U) = \int \dot N(E) \cdot T'(E,U) ~\mathrm{d}E ~+~ b ~=~ \dot N \otimes T'+b.
\label{eq:fitfunction}
\ee
The response function $T'(E,U)$ itself is a fivefold convolution of the transmission function $T_\mathrm{spec}$
of the spectrometer of MAC-E-Filter type, the energy loss function of the \belec s in the T$_2$ film $f_\mathrm{loss}$ \cite{aseev00}, 
the charge-up potential in the film $f_\mathrm{charge}$ \cite{bornschein03}, the backscattering 
function from the graphite (HOPG) substrate $f_{\rm back}$, and the energy dependence of the
detector efficiency $f_{\rm det}$ \cite{kraus05}:
\be
T'(E,U)=T_\mathrm{spec} \otimes f_\mathrm{loss} \otimes f_\mathrm{charge} \otimes f_\mathrm{back} \otimes f_\mathrm{det}.
\label{eq:tstrich}
\ee
The five functions $T_\mathrm{spec}$, $f_\mathrm{loss}$, $f_\mathrm{charge}$, $f_{\rm back}$, $f_{\rm det}$ are described in detail in the Mainz 
phase II analysis paper \cite{kraus05} and we copy the former analysis methods by even applying the same computer programs for these five functions
and for $T'(E,U)$.

For calculating the \bspec\ we use the final states distribution from the Mainz phase II analysis \cite{saenz00} including also the excitation of neighboring T$_2$ 
molecules during the \bdec\ and small shifts of higher excited electronic states in solid T$_2$ with respect to gaseous T$_2$ as described in the Mainz phase II analysis \cite{kraus05}. Although there are two slightly updated final states distributions \cite{doss06,doss08} the differences are so tiny that we still used the one 
\cite{saenz00}, which has been applied for the Mainz phase II analysis.

Fitting was done by the usual $\chi^2$ minimization method applying the program {\sc minuit} from CERN. Our systematic uncertainties we derived in the same
way as for the Mainz phase II analysis: For every parameter $p$ with systematic uncertainty $\Delta p$, e.g. the thickness of the T$_2$ film, we performed the whole 
fitting three times, with the parameter set to $p$, to $p-\Delta p$ and to $p+\Delta p$, respectively. The obtained variations in our observable of interest $\sin^2 \vartheta$
for a fixed mass squared of the fourth neutrino mass state \mtwonufour\ defined the systematic uncertainty to our observable 
$\pm 1 \sigma_{p, \mathrm{sys}}(\sin^2 \vartheta)$ by the parameter $p$.
Since the uncertainty $\Delta p$ of the parameter $p$, e.g. the uncertainty of the film thickness, usually differs among the six data sets we calculated the correct average by minimizing the $\chi^2$  for all six data sets together, as described in the Mainz phase II analysis \cite{kraus05}.

We briefly report on the systematic uncertainties, which we took into account in the same way as in the Mainz phase II neutrino mass search \cite{kraus05}:
\begin{enumerate}
\item{\bf Final states of the daughter molecule:} We use the 
calculation by Saenz et al. \cite{saenz00}, which was performed for 
gaseous T$_2$ with fully satisfactory precision compared to the additional uncertainties due to the solid state of the tritium source: 
In solid T$_2$ the 
excitation energy of higher excited final states shifts up slightly
with respect to the ground state of ($^3$HeT)$^+$. Similarly, the energy of excited electronic states in solid D$_2$ has been measured by inelastic electron scattering
to be shifted upwards with respect to gaseous T$_2$ \cite{aseev00}. 
The shift of the excited electronic levels of ($^3$HeT)$^+$ in solid $T_2$ has been
estimated by A.~Saenz \cite{saenz_private}. The shifts are larger for higher excited states.
Saenz found a correction of  0.8~eV for the second electronically excited state group and 
of 1.4~eV for the third one. Even higher excited states do not
play a significant role for this analysis. 
To be conservative, 
we consider the difference in the fit results on
\mtwonue\ with and without this correction fully as systematic uncertainty. 
\item{\bf Energy loss in the T$_2$ film:} The spectrum and the cross section of inelastic scattering
of the \belec s have been determined by the spectroscopy of $^\mathrm{83m}$Kr conversion electrons traversing thin D$_2$ films \cite{aseev00}.
The relative uncertainty of the cross section $\sigma_\mathrm{inelastic}$ was determined with a precision of 5.4\%, the relative uncertainty of the measurement of the film thickness  $\rho d$ by
laser ellipsometry was 3~\% averaged over all runs. Thus, 
the relative uncertainty of $\sigma_\mathrm{inelastic} \rho d$ amounts to 6.2~\%.
\item{\bf Hydrogen coverage of the T$_2$ film with time:}
By investigating the film thickness before and after the run as well as by analysing the time-dependent tritium $\beta$-spectra we found that
the T$_2$ had been covered during a run by a growing hydrogen film with a rate of 0.3 monolayers of H$_2$ per day. To account for possible systematic 
uncertainties of this description, we also rerun the analysis assuming a constant average film thickness. We consider the difference in the fit results on
\mtwonue\ of the two descriptions fully as systematic uncertainty. 
\item{\bf Neighbor excitation:} The prompt excitation of neighbors next 
to a decaying T$_2$ molecule has been estimated by Kolos in sudden
approximation \cite{kolos88}. The effect is due to the local relaxation
of the lattice following the sudden appearance of an ion. 
Kolos estimated an excitation probability of $P_{\rm ne}=0.059$ with a mean
excitation energy of $V_{\rm ne} =14.6$~eV. The latter number applies to the excitation
spectrum of free hydrogen molecules. We increased this 
number by the same 1.5~eV by which the energy loss spectrum of
electrons is shifted upwards in solid deuterium compared to gaseous hydrogen \cite{aseev00}. In the same
sense, the corresponding reduction of the total inelastic cross
section by 13\% \cite{aseev00} has been applied also to $P_{\rm ne}$ in the
analysis. Another reduction of $P_{\rm ne}$ by 11\% has
been accounted for the observed porosity of the quench-condensed films, yielding finally $P_{\rm ne}=0.046$. 
To be conservative,
we consider the difference in the fit results on \mtwonue\ of the two descriptions (Kolos' original values and our modified values for 
$P_{\rm ne}$ and for $V_{\rm ne}$)  fully as systematic uncertainty. 
It has been shown in \cite{kraus05} that the direct determination of $P_{\rm ne}$ from the Mainz tritium data gives similar results and similar systematic uncertainties, although the treatment of the systematics is very different due to a strong correlation of the derived $P_{\rm ne}$  with the energy loss.
\item{\bf Self-charging of T$_2$ film:} The quench-condensed T$_2$-films
charge up within 30~min to a constant critical field strength of $(62.6 \pm 4.0)$~MV/m 
\cite{bornschein03}. It results in a linearly
increasing shift of the starting potential of the \belec s
throughout the film, reaching about 2.5~V at the outer surface. In our analysis we
have assigned a conservative systematic uncertainty of $\pm 20\%$ to
that critical field strength as in the Mainz phase II analysis \cite{kraus05}.
\item{\bf Backscattering and energy dependence of detector efficiency:} Both effects are small
and can be accounted for by a linear correction factor $(1 + \alpha (E-qU))$ 
with the {\it surplus energy} $(E-qU)$ and with $q = -e$ the charge of the electron.
Depending on the run, $\alpha$ amounts to $6 \cdot 10^{-5}/$eV or $7 \cdot 10^{-5}/$~eV, respectively.
We assign a conservative uncertainty of $\Delta \alpha = \pm 4 \cdot 10^{-5}/$~eV.
\end{enumerate}

The final result on the light neutrino mass squared in equation (\ref{eq:mainz_phaseII}) was obtained by fitting the last 70~eV of the \bspec . In the sterile neutrino analysis we also used the last 70~eV of the measured $\beta$-spectra of the six runs for heavy neutrino masses 
$\mtwonufour \leq 1000~\mathrm{eV}^2$; for larger
\mnufour\ we extended the fit range proportional to \mnufour\ up to fitting the last 200 eV of the measured $\beta$-spectra for $\mtwonufour = 30000~\mathrm{eV}^2$.
Figure \ref{fig:uncertainties} presents the various systematic uncertainties as well as the total systematic uncertainty and the statistical error on
$\sin^2 \vartheta$ as a function of the squared mass of the fourth neutrino mass state \mtwonufour. It is clearly visible that the total systematic uncertainty is almost always  a bit larger than the statistical error bar. The former dominates the total error for $\mtwonufour \geq 3000~\mathrm{eV}^2$.
Since the data cover only the last 200 eV of the \bspec\ they are only sensitive to neutrino masses less than 200 eV. This explains the rise of the uncertainties on $\sin^2 \vartheta$ in figure \ref{fig:uncertainties} when \mtwonufour\ approaches $40000~\mathrm{eV}^2$ (the last fit was done for $\mnufour  = 180~\mathrm{eV}$).

\begin{figure}[t!]
    \centerline{\includegraphics[angle=0,width=0.9\textwidth]{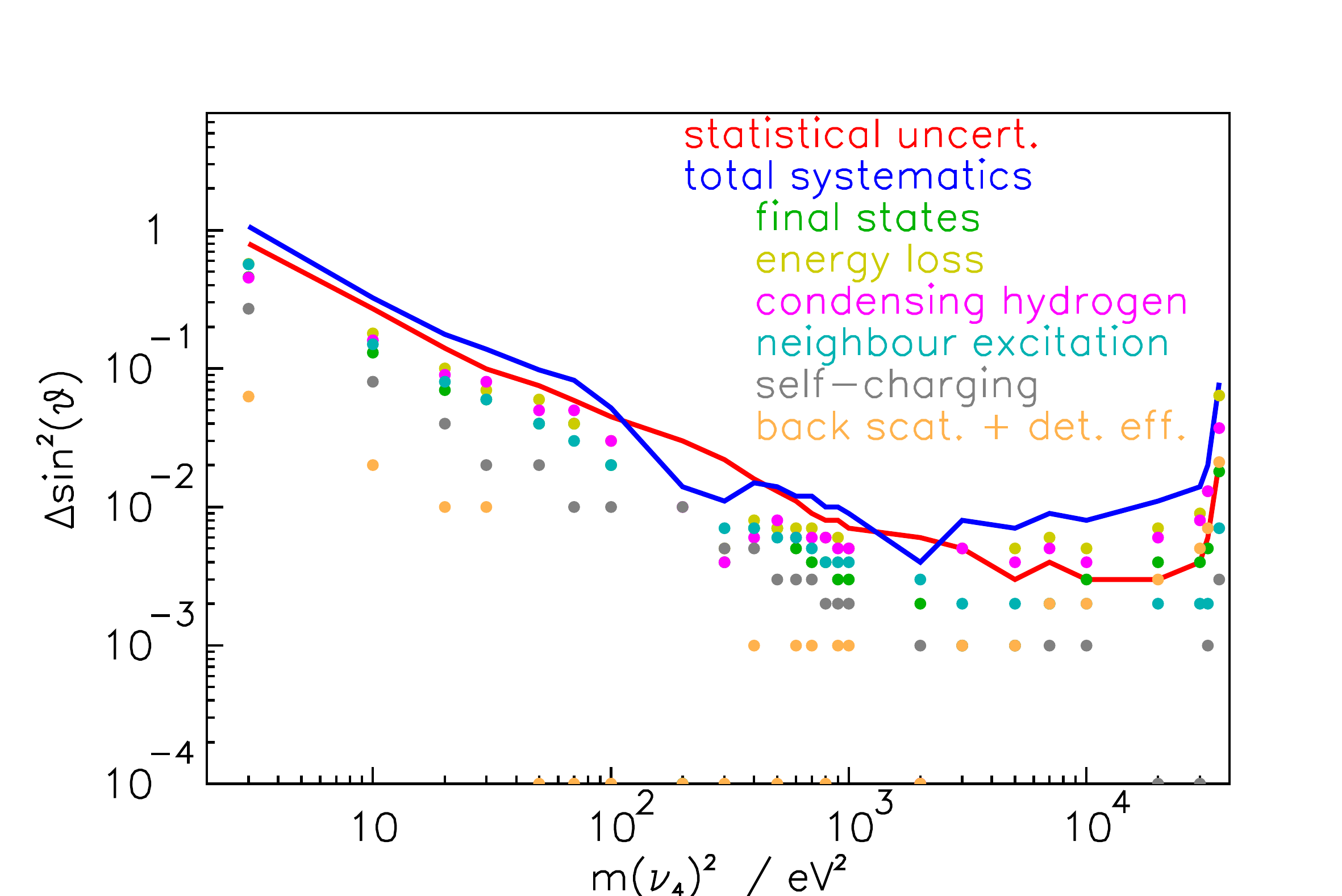}}
  \caption{Systematic and statistical uncertainties on the contribution $\sin^2 \vartheta$ of a fourth neutrino mass state $\nufour$ from the analysis of the Mainz phase II data
  as function of the squared mass \mtwonufour . The labels correspond to the description in section \ref{sec:analysis}. 
  The total systematic uncertainty is obtained as the
  square root of the sum of the squared individual systematic uncertainties, since the various systematic uncertainties are uncorrelated.} 
     \label{fig:uncertainties}
\end{figure}

\begin{figure}[h!]
    \centerline{\includegraphics[angle=0,width=0.9\textwidth]{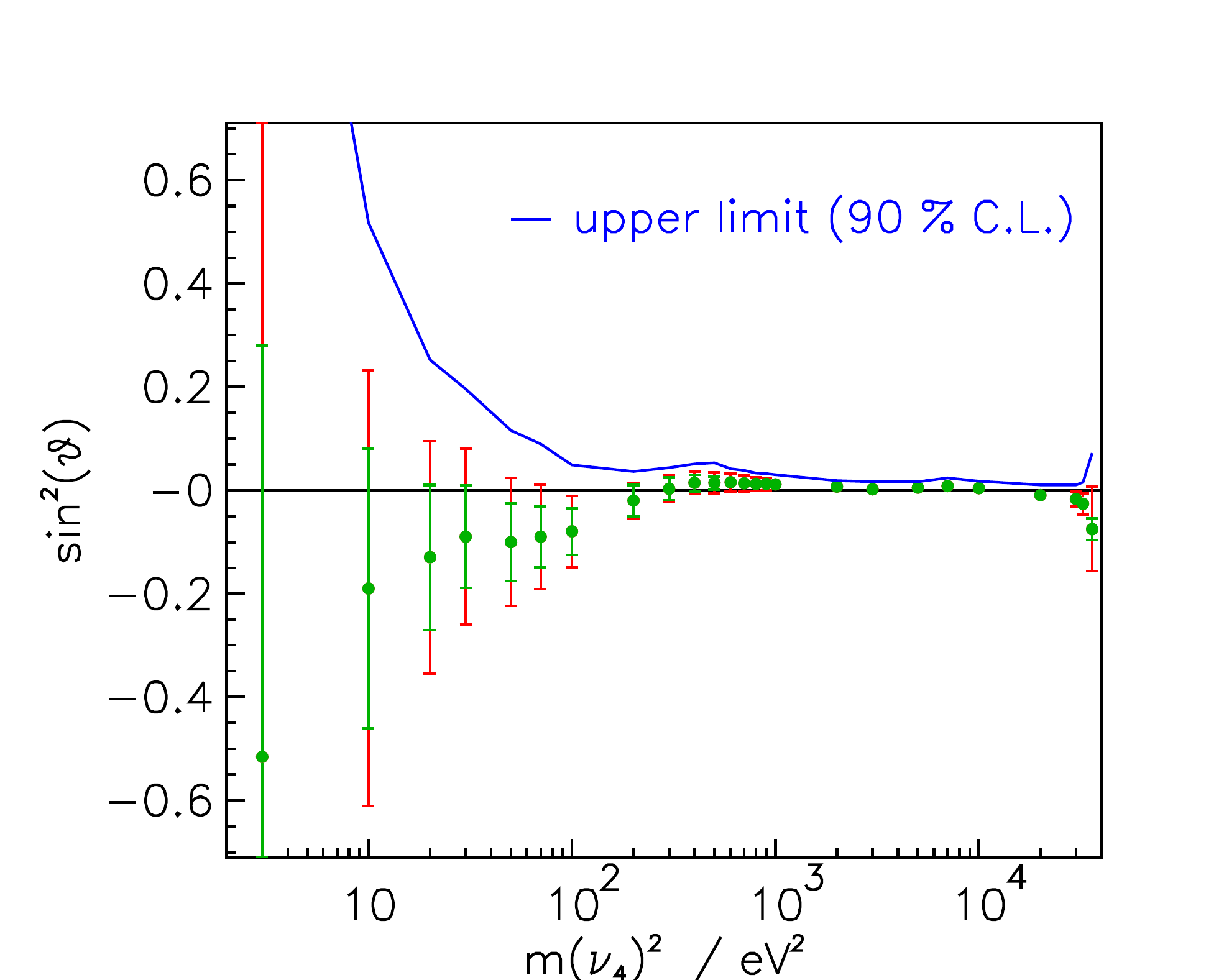}}
  \caption{Fit results on the contribution $\sin^2 \vartheta$ of a fourth neutrino mass state $\nufour$ from the analysis of the Mainz phase II data
  as function of the squared mass \mtwonufour .  
   The inner (green) error bars correspond to the statistical,
  the outer (red) to the total uncertainty. The blue line above the points with error bars gives the upper limit according to the Feldman-Cousins method  \cite{feldman98}
  with 90~\% C.L.}
    \label{fig:sin2theta_mnu4}
\end{figure}

Figure \ref{fig:sin2theta_mnu4} shows the fit results for the contribution $\sin^2 \vartheta$ as a function of the squared mass of the fourth neutrino mass state \mtwonufour.  For small \mtwonufour\ the sensitivity decreases due to lack of statistics. For larger squared masses \mtwonufour\ the variation in sensitivity  
is caused by the fact that the measurement points of the Mainz phase II runs are not equally distributed along the energy scale. 
For squared mass \mtwonufour\ approaching $40000~\mathrm{eV}^2$ the sensitivity decreases due to the smallness of the interval in which 
\mnufour\ influences the \bspec .
No indication of a  contribution of a fourth neutrino mass state is found, the contribution $\sin^2 \vartheta$ is compatible with zero for all squared 
masses \mtwonufour\ of the 
fourth neutrino mass state under investigation. The line above the fit points gives the corresponding upper limit according to the Feldman-Cousins method at 90~\% C.L. \cite{feldman98}.

Figure \ref{fig:reactor_neutrino_anomaly} shows the parameter space favored in order to explain the reactor neutrino anomaly by the mixing of a fourth neutrino mass state \cite{Men11} together with the limit on this fourth neutrino mass state from our analysis. The phase II data of the Mainz Neutrino Mass Experiment allow to exclude a small fraction of the favored parameter space at large $\Delta m^2$. 
Another limit on $\sin^2 (2 \vartheta )$ shown in figure \ref{fig:reactor_neutrino_anomaly} originates from a common analysis of data in the solar neutrino sector \cite{palazzo12}. 

Also shown are limits, which can be drawn under some assumptions from the search for neutrinoless double \bdec\ \cite{giunti12}:
If neutrinos are Majorana particles and if neutrinoless double \bdec\ dominantly works via non-zero neutrino masses then the lower limit on the half-life 
from neutrinoless double \bdec\ searches can be turned into upper limits on the effective neutrino mass \mbb . If we assume a 3+1 neutrino mixing scheme with three very light neutrinos and using the notation of equation (\ref{eq:two_nu_mixing}) we can connect \mbb\ with the mass of the fourth neutrino mass state \mnufour :
\be \label{eq:dbd}
  \mbb := | \sum_\mathrm{i=1}^4 U_\mathrm{ei}^2 \mnui | \approx |U_\mathrm{e4}^2 \mnufour| = \sin^2 \vartheta ~\mnufour \quad \mathrm{for~\mnui \approx 0,~i = 1,2,3}.
\ee

If the different neutrino mass states cannot be resolved in a single \bdec\ experiment, the measurement is sensitive to the 
so-called electron neutrino mass "\mnue" \cite{otten08,drexlin12}. Again under the assumption of a 3+1 neutrino mixing scheme with three very light neutrinos and using the notation of equation (\ref{eq:two_nu_mixing}) we can connect \mnue\ with the mass of the fourth neutrino mass state \mnufour :
\be \label{eq:mnue}
  \mtwonue: =  \sum_\mathrm{i=1}^4 |U_\mathrm{ei}|^2 \mtwonui  = \sin^2 \vartheta ~\mtwonufour \quad \mathrm{for~\mnui \approx 0,~i = 1,2,3}.
\ee
Using this equation and the neutrino mass limit from the Mainz phase II data of equation (\ref{eq:mainz_phaseII_limit}) 
we can extract a sensitivity estimate of the Mainz phase II data for a fourth neutrino mass state, which is shown in figure 
\ref{fig:reactor_neutrino_anomaly} as thin black line. For low masses, where the simplification of equation (\ref{eq:mnue}) holds, the sensitivity estimate
agrees rather well with the limits obtained in the detailed analysis described in this paper. For larger squared neutrino masses \mtwonufour\ the real
limit becomes less sensitive. 

We do not plot the limit on the admixture of a fourth heavy neutrino obtained in a recent analysis by part of the Troitsk collaboration \cite{troitsk12}. This limit
is significantly more stringent than the one from our analysis although both are based on similarly sensitive measurements of the tritium $\beta$-spectrum. We cannot follow the arguments put forth by the authors of reference \cite{troitsk12} who claim that the systematic uncertainties do not need to be considered. 
In their recent standard neutrino mass analysis \cite{aseev11} the Troitsk collaboration obtained a systematic uncertainty on \mtwonue\ about as large as the corresponding statistical uncertainty, similar to the Mainz results stated in equation (\ref{eq:mainz_phaseII}).
Figure \ref{fig:uncertainties} clearly shows -- at least for the Mainz data -- that also for the search for a contribution of a fourth neutrino mass state systematic uncertainties 
are indeed very significant. We do not see how the new Troitsk analysis can give a constraint on $\sin^2 \vartheta$ at $\mnufour = 2$~eV at 95~\% C.L., which was the limit of the standard neutrino analysis from Troitsk \cite{aseev11}.

\begin{figure}[t!]
  \centerline{\includegraphics[angle=0,width=0.9\textwidth]{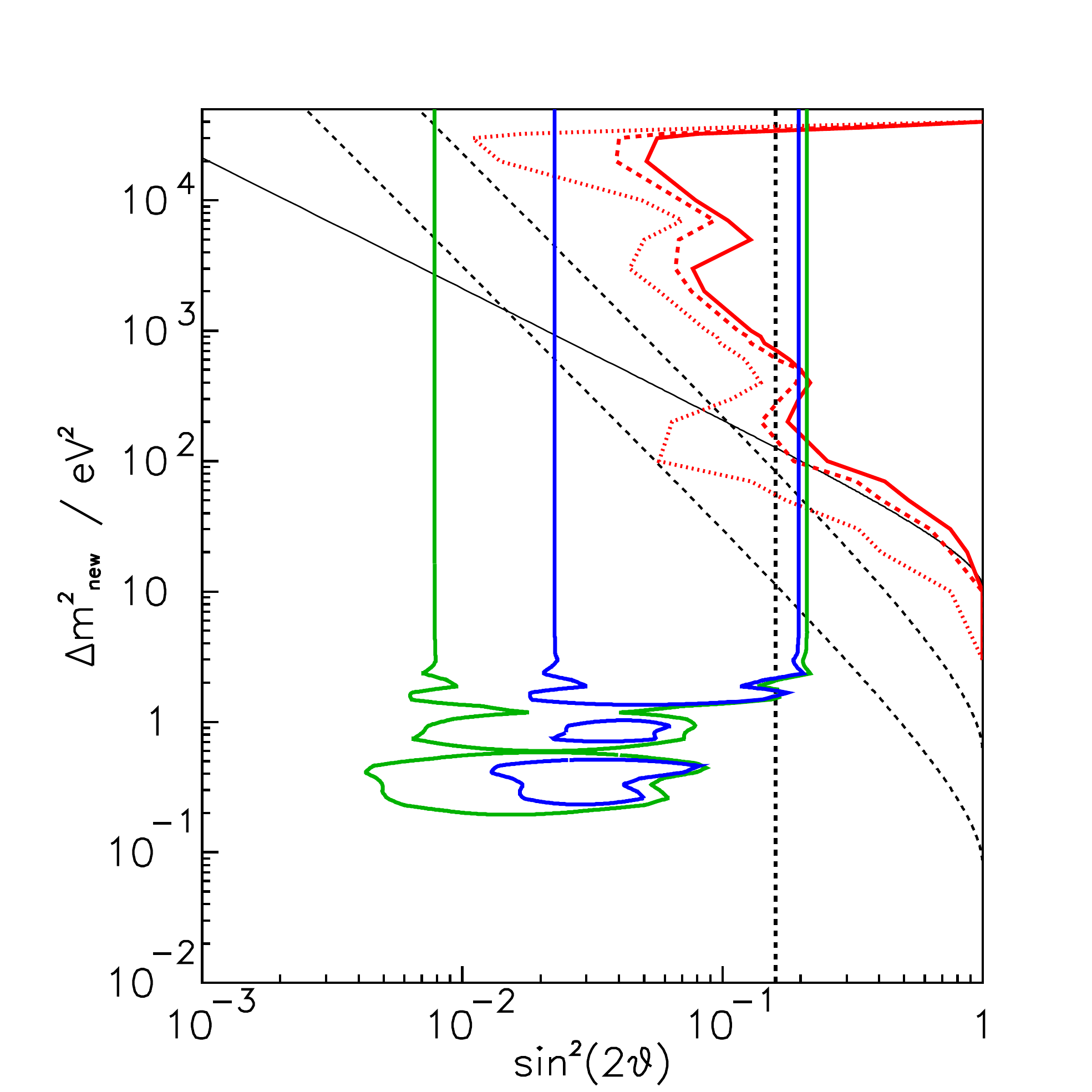}}
  \caption{Favored region of the so-called {\it reactor neutrino anomaly} at 90~\% C.L. in blue and at 95~\% C.L. in green. The data are shown as function of the mixing 
  of the fourth neutrino mass state to the electron neutrino $\sin^2 (2 \vartheta)$ and the squared mass difference  
  of the fourth neutrino mass state to the light neutrino mass states $\Delta m^2_\mathrm{new}$  (courtesy of T. Lasserre).
  The red curves represent the limits from this analysis at 68~\% C.L. (dotted), 90~\% C.L. (dashed) and 95~\% C.L. (solid), respectively. 
  The parameter regions right of  the curves are excluded. Here we neglect a possible non-zero value
  of the light neutrino mass states and show the limits for $\mtwonufour = \Delta m^2_\mathrm{new}$. 
  The black diagonal dashed lines are limits with 90~\% C.L. \cite{giunti12} derived with equation (\ref{eq:dbd}) from the search for neutrinoless double \bdec\ by the EXO-200 collaboration \cite{exo12} under the assumption that in a 3+1 neutrino mixing scenario the 3 light neutrino masses can be neglected; the two lines mark the range given by different nuclear matrix elements. The vertical black dashed line represents an upper limit on $\sin^2 2 \vartheta$ with 90~\% C.L. derived from a common analysis of neutrino oscillation data in  the solar neutrino sector \cite{palazzo12}. The black solid line represents the sensitivity estimate of the Mainz neutrino experiment from equation (\ref{eq:mnue}).}
    \label{fig:reactor_neutrino_anomaly}
\end{figure}

\section{Conclusion and outlook}

Our re-analysis of the phase II data of the Mainz Neutrino Mass Experiment with regard to a potential contribution of a fourth neutrino mass state 
to the electron neutrino does not give any hint for the existence of such a state. The contribution $\sin^2 \vartheta$ is compatible with zero for all squared 
masses of the fourth neutrino mass state under investigation (3~eV$^2 \leq \mtwonufour \leq 36400$~eV$^2$). 
The Mainz data constrain a small fraction of the parameter space 
 for such a fourth neutrino mass state favored by the attempt to explain the reactor neutrino anomaly and other indications. 

The Karlsruhe Tritium Neutrino experiment KATRIN \cite{KAT04} will investigate the endpoint region of the tritium \bdec\  with much higher 
statistics, better energy resolution and much smaller systematic uncertainties. The KATRIN experiment will reach a factor 10 higher sensitivity on the electron neutrino mass of 200~meV compared to the
sensitivity of the Mainz Neutrino Mass Experiment as reported in equation (\ref{eq:mainz_phaseII_limit}).
The data from the KATRIN experiment will also allow to investigate
the potential contribution of a fourth neutrino mass state to the electron neutrino with a sensitivity \cite{riis11,formaggio11,Esmaili12}
covering the whole favored region of the {\it reactor neutrino anomaly}.

A fourth neutrino mass state with a mass of a few keV acting as Warm Dark Matter is another possibility, which derives its motivation from recent efforts of explaining the structure of the universe at 
galactic and super-galactic scales ({\it missing satellite galaxy problem}) \cite{davega11a}. 
Such a neutrino might also be investigated by \bdec\ studies \cite{davega11b}.

\section*{Acknowledgement}

We would like to thank our colleagues from the former Mainz Neutrino Mass Experiment, especially Ernst Otten, for helpful discussions. We would like
to thank Robert Shrock, who asked us to do this analysis long before the {\it reactor neutrino anomaly} was announced. We would like to thank Thierry Lasserre for  providing us with the 2-dim $\chi^2$ distribution of the {\it reactor neutrino anomaly} analysis of reference \cite{Men11}.
We would like to thank Werner Rodejohann who pointed us to similar limits on a fourth neutrino mass state, which can be drawn by data from neutrinoless double \bdec , which we cite from reference \cite{giunti12}.

\end{document}